\documentclass[aps,prl,floatfix,amsmath,amssymb,reprint]{revtex4-2}
\usepackage{graphicx}
\usepackage{color}
\usepackage{microtype}
\usepackage{soul}

\usepackage[colorlinks=true, allcolors=blue]{hyperref}

\def\equationautorefname#1#2\null{Eq.#1(#2\null)}

\newcommand{\dd}{\mathrm{d}}
\newcommand{\BRLS}{BR\,LS}
\newcommand{\HsNH}{\hat{H}_s^{\mathrm{NH}}}
\newcommand{\HsNHd}{\hat{H}_s^{\mathrm{NH}\dagger}}

\usepackage[export]{adjustbox}
\usepackage[FIGBOTCAP]{subfigure}

\begin{document}
\title{Memory loss is contagious in open quantum systems}

\author{Anael Ben-Asher}
\affiliation{Departamento de Física Teórica de la Materia Condensada and Condensed Matter Physics Center (IFIMAC), Universidad Autónoma de Madrid, E28049 Madrid, Spain}

\author{Antonio I. Fern\'andez-Dom\'inguez}
\affiliation{Departamento de Física Teórica de la Materia Condensada and Condensed Matter Physics Center (IFIMAC), Universidad Autónoma de Madrid, E28049 Madrid, Spain}

\author{Johannes Feist}
\affiliation{Departamento de Física Teórica de la Materia Condensada and Condensed Matter Physics Center (IFIMAC), Universidad Autónoma de Madrid, E28049 Madrid, Spain}

\begin{abstract}
Memoryless (Markovian) system-bath interactions are of fundamental interest in physics. While typically, the absence of memory originates from the characteristics of the bath, here we demonstrate that it can result from the system becoming lossy due to the Markovian interaction with a second bath. This uncovers an interesting interplay between independent baths and suggests that Markovianity is ``contagious'', i.e., it can be transferred from one bath to another through the system with which they both interact. We introduce a Bloch-Redfield-inspired approach that accounts for this distinct origin of Markovianity and uniquely combines non-Hermitian Hamiltonian formalism with master equations. This method significantly improves the description of the interaction between a lossy system (associated with a Lindblad master equation) and a non-Markovian bath, reducing the computational demands of complex system-bath setups across various fields. 
\end{abstract}
\maketitle

No physical system is ever fully isolated from its environment, and the possibility of exchanging energy or particles with the baths in the environment induces dissipation. This is a fundamental aspect of open quantum systems~\cite{breuer2002theory,de2017dynamics}, ubiquitous in nature and technology.
System-bath interactions are typically classified as either Markovian (memoryless) or non-Markovian.
Markovianity,  which refers to the evolution of a system that depends only on its present state, holds great importance not only for characterizing many processes but also  as an efficient approximation that reduces conceptual and computational complexity~\cite{de2017dynamics,breuer2002theory,dhahri2019open,noe2019markov,vicentini2019variational}.
In open quantum systems, the bath autocorrelation function  $C(t)$ describes the memory the bath retains at time $\tau+t>\tau$ of its interaction with the system at time $\tau$~\cite{de2017dynamics,breuer2002theory}. Therefore, a bath is considered Markovian when $C(t)$ decays to zero faster than any other timescale of the problem, a property determined solely by the internal structure and dynamics of the bath. 

\begin{figure}[tb]
   \includegraphics[width=0.45\columnwidth, angle=0,scale=1,
draft=false,clip=true,keepaspectratio=true,valign=t]{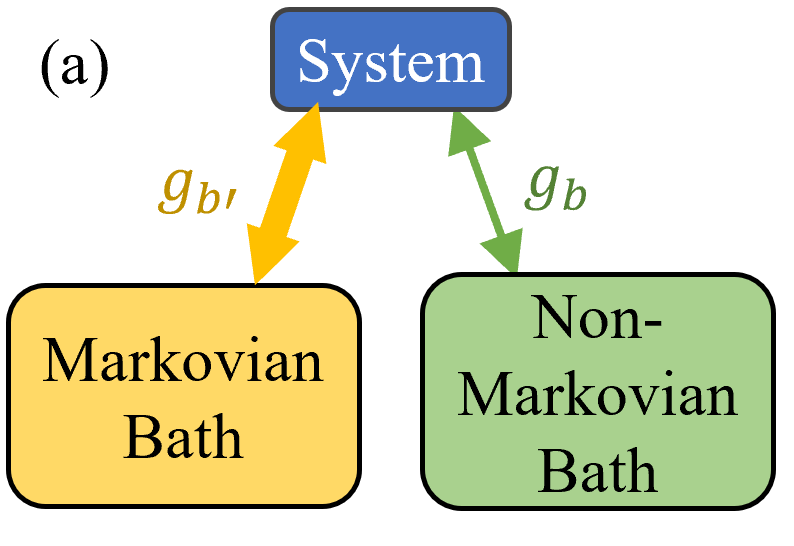}
   \includegraphics[width=0.45\columnwidth, angle=0,scale=1,
draft=false,clip=true,keepaspectratio=true,valign=t]{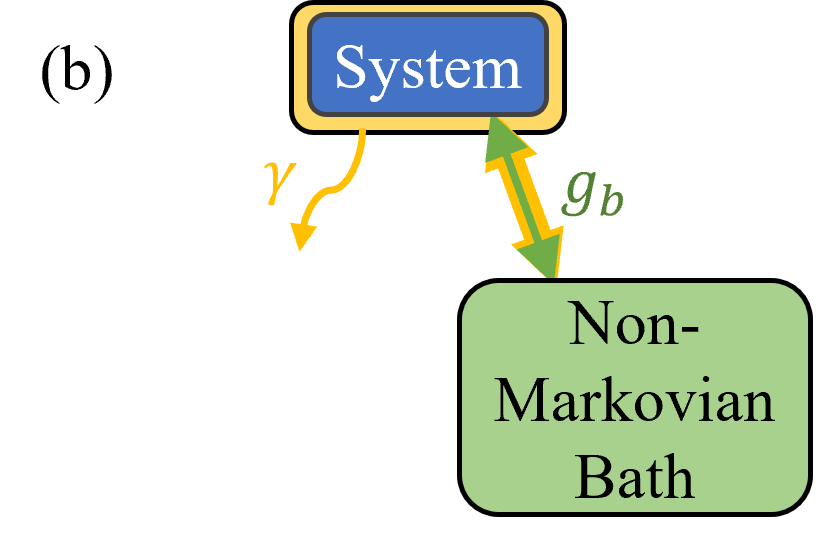}

\vspace{0.4cm}

   \includegraphics[width=0.9\columnwidth, angle=0,scale=1,
draft=false,clip=true,keepaspectratio=true]{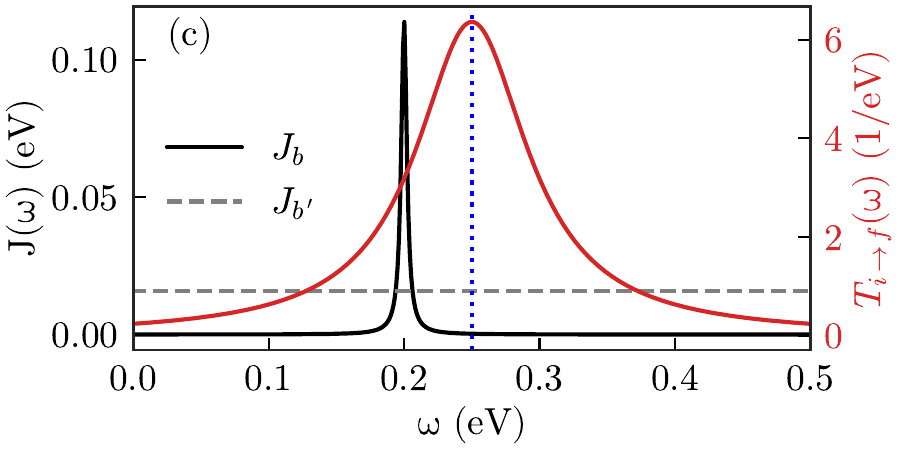}
\caption[]{(a), (b) Two schemes describing a lossy system interacting with a non-Markovian bath. (c)  Spectral densities of a non-Markovian bath (in black) and a Markovian bath (in grey) and an example of a transition spectrum between two lossy states ($T_{i\to f}(\omega)$ in red).}
\label{fig:1}
\end{figure}

In this Letter, we demonstrate that Markovianity can also arise due to the properties of the system rather than solely from the bath. Specifically, it can stem from the dissipative character of the system, which results in a faster memory loss of its interaction with a non-Markovian bath. The origin of the dissipation in the system is its interaction with another bath that is Markovian  (as sketched in \autoref{fig:1}) and the interplay between the two competing baths results in this novel type of memoryless interaction. When two independent baths interact with the system through coupling operators that do not commute with each other and with the system Hamiltonian, they cannot be described by a single effective bath whose effect is merely the sum of the individual effects of each bath~\cite{kolodynski2018adding,chan2014quantum,gribben2022exact}. Accounting for their combined effect usually requires computationally demanding approaches such as tensor network methods~\cite{gribben2022exact} or approximate methods such as the polaron transformation~\cite{bundgaard2021non} or the reaction coordinate framework~\cite{maguire2019environmental,mcconnell2019electron}.
We alternatively suggest that when one of the baths is memoryless, its effect on the system's interaction with the other bath can be captured by a master equation derived within the interaction picture of the effective non-Hermitian (NH) Hamiltonian~\cite{roccati2022non} encoding the dissipation of the memoryless bath. This unique NH framework enables us to express the combined effect of the two baths as the dissipation induced by the Markovian bath plus the interaction of the system with the non-Markovian bath, dressed by the former dissipation. Furthermore, it sheds light on the interplay between the memory effects of the baths, showing that due to this dressing, the originally non-Markovian interaction can become Markovian. This implies that, in some sense, Markovianity is ``contagious'' and can be transferred from one bath to another through the system with which they both interact.

We consider a quantum system coupled to two independent bosonic baths, one non-Markovian and the other Markovian, with the coupling strengths $g_b$ and $g_{b'}$ as shown in \autoref{fig:1}(a). This setup is described by the Hamiltonian:
\begin{equation}
\hat{H}_{s+b+b'}=\hat{H}_s+\hat{H}_{b}+\hat{H}_{b'}+g_b\hat{V}_s\hat{U}_b+g_{b'}\hat{V}_s'\hat{U}_{b'},\label{fullH}
\end{equation}
where $\hat{H}_s,\hat{H}_b,\hat{H}_{b'}$ are the system and the two baths Hamiltonians;
$\hat{V}_s, \hat{V}_s'$ are the system operators that couple to each bath and obey either $[\hat{V}_s,\hat{V}_s']\neq0$ or $[\hat{H}_s,\hat{V}_s']\neq0$~\cite{kolodynski2018adding}; and $\hat{U}_b=\int_0^\infty d\omega \sqrt{J_{b}(\omega)/g_b^2}(b_\omega+b_\omega^\dagger)$ and $\hat{U}_{b'}=\int_0^\infty d\omega \sqrt{J_{b'}(\omega)/g_{b'}^2}(b_\omega'+b_\omega'^\dagger)$ are the baths operators which couple them to the system, where $b_\omega, b_\omega'$ ($b_\omega^\dagger, b_\omega'^\dagger$) are the bath bosonic annihilation (creation) operators, and $J_b(\omega), J_{b'}(\omega)$ are the baths spectral densities, related to their autocorrelation functions when the temperature is zero by $J_b(\omega)=\int_0^\infty dt C_b(t)e^{i\omega t}$ and $J_{b'}(\omega)=\int_0^\infty dt C_{b'}(t)e^{i\omega t}$. While $J_b(\omega)$ is structured and sharp in energy, $J_{b'}(\omega)$ is unstructured and even constant, as is exemplified in \autoref{fig:1}(c). The width of the system-bath arrows in \autoref{fig:1}(a)--(b) indicates the Markovian (non-Markovian) character of the interaction, representing the associated broad (sharp) energy spectrum.

First, we use the memoryless character of the Markovian bath and treat its influence in accordance with standard open quantum systems theory using the Lindblad formalism~\cite{breuer2002theory,lindblad1976generators}. This results in the master equation~\cite{breuer2002theory,lindblad1976generators}, 
\begin{equation}
\frac{\dd\rho_{s+b}}{\dd t}=-\frac{i}{\hbar}[\hat{H}_s+\hat{H}_b+g_b\hat{V}_s\hat{U}_b, \rho_{s+b}]+\sum_{A'}\gamma_{A'}\hat{L}_{\hat{A}'}[\rho_{s+b}]\label{LinMEQ}
\end{equation}
associated with the setup of a lossy system interacting with a non-Markovian bath presented in \autoref{fig:1}(b) using the total loss rate $\gamma$. 
Here, $\rho_{s+b}$ is the density matrix that includes the degrees of freedom of the system and the non-Markovian bath. The system's interaction with the Markovian bath is encoded by the Lindblad superoperators: $\hat{L}_{\hat{A}'}[\rho] = {\hat{A}'^\dagger} \rho {\hat{A}'} - \frac{1}{2}\{{\hat{A}'^\dagger} {\hat{A}'}, \rho\}$, where $\hat{A}'$ are the jump operators associated with the loss rates $\gamma_{A'}$ and originating from $\hat{V}_s'$ and $J_{b'}(\omega)$~\cite{breuer2002theory}. 
The anticommutator in $\hat{L}_{\hat{A}'}[\rho]$ describes decaying population in the coherent evolution of the system due to dissipation to the Markovian bath, while the other, ``quantum jump'' or ``refilling'' term describes the reappearance of this population. Such lossy dynamics cause decoherence, leading to the system's energy levels not being well-defined but instead having a bandwidth. 

To account for the effect of the Markovian bath in the system's interaction with the non-Markovian bath, we rely on the effective NH Hamiltonian~\cite{roccati2022non}, arising from the coherent terms of the Lindblad superoperator shown in \autoref{LinMEQ} and given by $\HsNH = \hat{H}_s-\frac{i}{2}\sum_{A'}\gamma_{A'} \hat{A}'^\dagger\hat{A}'$. Effective NH Hamiltonians are commonly used as an alternative formalism to the Lindblad master equation~\cite{roccati2022non,ben2023non,felicetti2020photoprotecting,fabri2024coupling}, valid when the jump operators $\hat{A}'$ transfer the population out of the manifold of interest and therefore the ``refilling'' term of the Lindblad superoperator can be omitted. We, however, use the NH Hamiltonian to simplify the treatment of the non-Markovian bath even when it cannot describe all the system's dynamics and serve as an alternative formalism. The NH Hamiltonian captures the lossy character induced by the Markovian bath, thus incorporating its effect into the system as described in \autoref{fig:1}. This facilitates a direct treatment of the interaction between the resulting lossy system and the non-Markovian bath.

We draw inspiration from the Bloch-Redfield (BR) theory~\cite{wangsness1953dynamical,redfield1955nuclear}, designed for the perturbative and Markovian regime of the system-bath interaction and yielding under a secular approximation the Lindblad description~\cite{breuer2002theory},  to derive a memoryless master equation for the system density matrix $\rho_s(t)$ from the interaction between the lossy system (incorporating the influence of the Markovian bath) and the non-Markovian bath. 
The full derivation of this BR for lossy systems (\BRLS) approach is given in the Supplemental Material (SM). 
While the standard BR approach assumes a Markovian bath, \BRLS{} takes into account the broadening of the lossy energy levels and considers it as the origin of the memoryless interaction with a non-Markovian bath. For that, we utilize the eigenstates of $\HsNH$ and $\HsNHd$, denoted as $|i)$ and $|j^*)$ and associated with the complex eigenvalues $\omega_i-i\frac{\Gamma_i}{2}$ and $\omega_j+i\frac{\Gamma_j}{2}$, respectively.  The real part $\omega_{i/j}$ encodes the states' energy position, while the imaginary part $\Gamma_{i/j}$ encodes the loss rate of the states, also associated with their linewidths~\cite{moiseyev2011non}. 
The Markovian assumption is then applied in a frame that describes the system's loss of coherence in addition to the dynamics arising from the interaction with the non-Markovian bath, assuming $(i|\rho_s(t-\tau)|j^*)\approx(i|\rho_s(t)|j^*)e^{i(\omega_i-\omega_j)\tau}$ (see SM) rather than ${\rho}_{s}^I(t-\tau)\approx{\rho}_{s}^I(t)$  where ${\rho}_{s}^I(t)=e^{i\HsNH t/\hbar}{\rho}_s(t)e^{-i\HsNHd  t/\hbar}$. This approximation facilitates assuming  Markovian system-bath interaction due to the system loss without sacrificing the memory of the system's internal dynamics.
Note that for eigenstates of a NH Hamiltonian, the notation $|\dots)$, $(\dots|$ rather than $|\dots\rangle$, $\langle\dots|$ is used to describe the right and left eigenstates~\cite{moiseyev2011non}. 

The resulting master equation, including the effects of both baths, reads as
\begin{eqnarray}
\frac{\dd{\rho_s}(t)}{\dd t}=-\frac{i}{\hbar}[\hat{H}_s, \rho_{s}(t)]+\sum_{A'}\gamma_{A'}\hat{L}_{\hat{A}'}[\rho_{s}]+\\\nonumber\sum_{a,b,c,d}R_{abcd}|a)(c|\rho_s(t)|d^*)(b^*|,
\end{eqnarray} 
where the Lindblad superoperators are present, also including the ``refilling" terms omitted in the derivation of the \BRLS{} tensor. 
The first term describes the system's internal dynamics, the second term encodes the dissipation induced by the memoryless bath, and the third term presents the influence of the non-Markovian bath dressed by the former dissipation and given by the \BRLS{} tensor,
\begin{eqnarray}
\label{tensor}
R_{abcd} = \delta_{bd} \sum_q F_{cdq}\tilde{V}_{aq}\tilde{V}_{qc} + \delta_{ac} \sum_q F_{dcq}^* \tilde{V}^*_{dq} \tilde{V}_{qb}^*\\\nonumber
 -(F_{cda}+F_{dcb}^*)\tilde{V}_{ac}\tilde{V}_{db}^*, 
\end{eqnarray}
where $F_{jsq}=\int_0^\infty d\tau C_b(\tau)e^{-\frac{1}{2}(\Gamma_q+\Gamma_s)\tau/\hbar}e^{-i(\omega_q-\omega_j)\tau}$ is the unilateral Laplace transform of  $C_b(t)$, 
$\tilde{V}_{ij}=(i|\hat{V}_s|j)$ and $\tilde{V}_{ij}^*=(i^*|\hat{V}_{s}|j^*)$. Importantly, \autoref{tensor} reproduces the standard BR tensor~\cite{breuer2002theory,peskin2023quantum} when neglecting the losses induced by the other bath. While the BR approach is accurate when $C_b(\tau)$ decays rapidly, as happens for Markovian baths, \BRLS{} only requires rapid decay of $C_b(\tau)e^{-\frac{1}{2}(\Gamma_q+\Gamma_s)\tau/\hbar}$, which takes into account the losses of the system. 
This sheds light on the interplay between the two baths in inducing memoryless dynamics, showing that when the Markovian bath causes sufficiently large losses to the system, the influence of the non-Markovian bath can be described by a Markovian term using $R_{abcd}$. 

The \BRLS{} approach introduced above is relevant for any system featuring losses and interacting with an external bath independently from the Markovian bath responsible for them. 
Notably,  \BRLS{} yields accurate results even in cases where the effective NH Hamiltonian formalism, essential for its derivation, cannot describe the whole dynamics and the complete Lindblad master equation is needed (see SM). 
However, it is not valid when the losses are larger than the system's energies and the Lindblad formalism may introduce artificial pumping of the ground state~\cite{lednev2024lindblad}. 

We apply the \BRLS{} for a cavity-quantum electrodynamical (QED) setup in which quantum emitters (QEs) are strongly coupled to a cavity light mode. These setups attract great attention, offering the manipulation of quantum properties of both light and matter~\cite{galego2016suppressing,zhong2017energy,garcia2021manipulating,raimond2001manipulating,reiserer2015cavity,wang2019turning}.
They are lossy systems due to imperfect reflection and absorption of the cavity that enables photon leakage to a Markovian bath. 
In particular, metallic nanocavities or nanoresonators, which can achieve the strong-coupling regime in the QE-cavity interaction, feature large decay rates~\cite{torma2014strong,chikkaraddy2016single,zhang2017sub,kongsuwan2018suppressed,baumberg2019extreme}.
Several types of QEs used in cavity QED setups, such as molecular J- and H- aggregates~\cite{spano2015optical,herrera2016cavity,saez2018organic} or quantum dots~\cite{grange2017reducing,peng2023polaritonic}, can be described as two-level QEs coupled to structured phononic baths.
Then, the system Hamiltonian is given by the NH Tavis-Cummings (TC) model~\cite{tavis1968exact}:
\begin{equation}
\HsNH = \tilde{\omega}_c a^\dagger a+\sum_j^N \left[\tilde{\omega}_e\sigma^j_+\sigma^j_-+\frac{g_{ec}}{\sqrt{N}}(a^\dagger \sigma^j_-+\sigma^j_+ a)\right] \label{Hm}
\end{equation}
obtained from the Lindblad superoperators $\gamma_c\hat{L}_{\hat{A}'=a}$ and $\gamma_e\hat{L}_{\hat{A}'=\sigma^j_{-}}$, and the coupling operator of each $j$-th QE to its non-Markovian bath is $\hat{V}^j_s=\sigma^j_+\sigma^j_-$.
Here, $a$ ($a^\dagger$) and $\sigma^j_-$ ($\sigma^j_+$) are the annihilation (creation) operators for the cavity mode and the $j$-th two-level QE, respectively. The complex energies of the cavity mode and the QEs are denoted by $\tilde{\omega}_c=\omega_c-\frac{i}{2}\gamma_c$ and $\tilde{\omega}_e=\omega_e-\frac{i}{2}\gamma_e$, respectively, and the coupling strength between the cavity mode and each QE is $g_{ec}/\sqrt{N}$.
In the following, we use the population transfer between the eigenstates of the NH Hamiltonian in \autoref{Hm}, which is zero for $g_b=0$ (see SM), to isolate and analyze only the dynamics induced by the non-Markovian bath.  
When $g_{ec}>\frac{|\gamma_c-\gamma_e|}{4}$ and $\omega_e=\omega_c$, $\HsNH$ has two hybrid light-matter eigenstates, known as polaritons, with complex energies ${\omega}_e\pm g_{ec}-\frac{i}{2}\Gamma_p$ where  $\Gamma_p=\frac{\gamma_e+\gamma_c}{2}$ is their loss rate.
The lower-energy state is typically called the lower polariton (LP), while the higher-energy state is known as the upper polariton (UP). 
The other states arising when $N\geq 2$ are pure excitonic states with the complex energy $\tilde{\omega}_e$, usually termed as dark states (DS)~\cite{del2015quantum}. 

\begin{figure}[t]
\begin{center}
   \includegraphics[width=\columnwidth, angle=0,scale=1,
draft=false,clip=true,keepaspectratio=true]{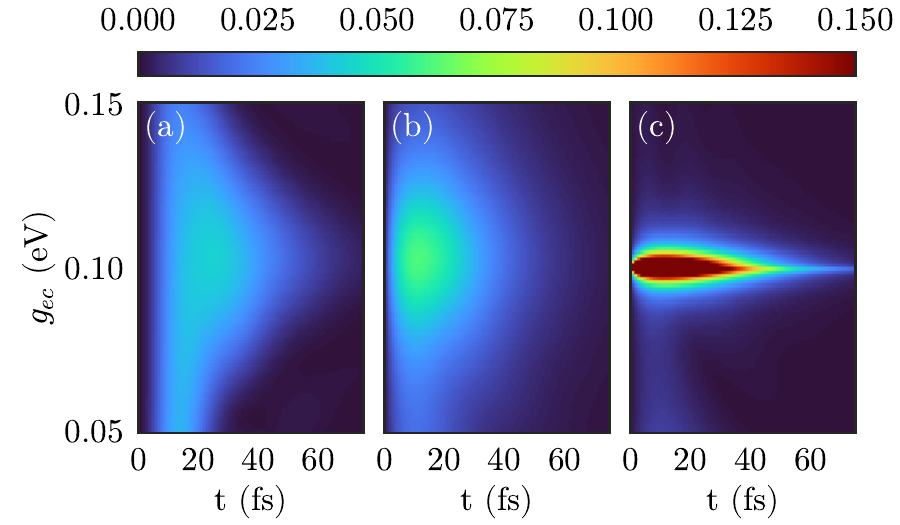}

\caption[]{Bath-driven population transfer to the LP of \autoref{Hm} ($N=1$) obtained by (a) the numerically exact calculation, (b) the introduced \BRLS{} approach, and (c) the standard BR approach, when only the UP is populated at $t=0$.
}
\label{fig:2}
\end{center}
\end{figure}

As a test case, we consider a single QE ($N=1$ in  \autoref{Hm}) with an, in principle, non-Markovian phononic bath whose spectral density is plotted in black in \autoref{fig:1}(c) and given by
\begin{eqnarray}
J_b(\omega)=\Theta(\omega)\frac{2g_b^2}{\pi}\frac{\kappa\omega\omega_b}{(\omega^2-\omega_b^2)^2+\kappa^2\omega^2},\label{bathspec}
\end{eqnarray}
where $\Theta(\omega)$ is the Heaviside function that is zero for $\omega<0$, $\omega_b=0.2$ eV, $\kappa=5$ meV and $g_b= 0.03$ eV. No thermal population of the bath is considered. 
This test case can be solved by the numerically exact discretization method~\cite{lednev2024lindblad}, enabling the benchmark of \BRLS{}  and the analysis of the dynamics' Markovian character.
In \autoref{fig:2}, we compare the population of the LP calculated by the \BRLS{} and the numerically exact approaches using QuTiP\cite{johansson2012qutip}, as a function of the Rabi splitting $\omega_{\mathrm{UP}}-\omega_{\mathrm{LP}}\approx2 g_{ec}$ and the time $t$, where the initial state is the UP, $\gamma_c=0.1$~eV, $\gamma_e=0.1$~meV and $\omega_c=\omega_e$. 
In addition, we present in \autoref{fig:2}(c) the population transfer obtained by the BR method that treats the phononic bath independently from the system losses~\cite{del2015quantum}. 
As BR is incapable of treating non-Markovian baths and does not account for the interplay between the two baths, the results that it yields (\autoref{fig:2}(c))  significantly differ from the exact results (\autoref{fig:2}(a)). 
However, the \BRLS{} (\autoref{fig:2}(b)) obtains much better agreement, particularly for times longer than the lifetime of the states, i.e., $t>\tau_p=\frac{1}{\Gamma_p}\approx13$ fs, verifying the memory loss of the interaction with the non-Markovian bath due to the system loss as assumed within the \BRLS. 
The performance of the different approaches in computing the elements of the system density matrix, as well as an analysis of the Markovianity of the full dynamics based on a quantity measuring backflow of information into the system~\cite{breuer2009measure}, are presented in the SM\@. These confirm that the introduction of losses in the system increases the Markovianity of its dynamics. 

\begin{figure}[t]
\begin{center}
\includegraphics[width=\columnwidth, angle=0,scale=1,
draft=false,clip=true,keepaspectratio=true,valign=t]{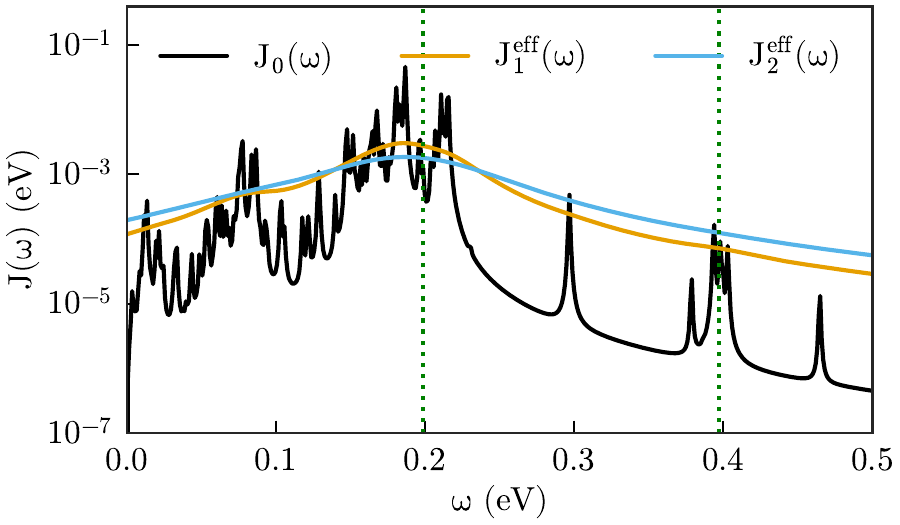}
\caption[]{A structured spectral density J$_0(\omega)$ modeling the vibrational structure of a J-aggregated dye (in black) and the effective spectral densities (defined in \autoref{Jeff}) for the transitions UP$\to$DS and DS$\to$LP (J$_1^{\mathrm{eff}}(\omega)$) and UP$\to$LP (J$_2^{\mathrm{eff}}(\omega)$).} \label{fig:3}
\end{center}
\end{figure}

The numerically exact results shown in \autoref{fig:2}(a) present bath-driven population transfer over a larger range of Rabi splittings than would be expected from inspecting the bandwidth of the bath spectral density $J_b(\omega)$ presented in \autoref{fig:1}(c). 
We use the \BRLS{} approach to show that this broadening stems from the broad linewidth of the energy-dependent transition spectrum, given by $T_{i\to f}(\omega)=\frac{1}{\pi}\frac{\frac{1}{2}(\Gamma_i+\Gamma_f)}{(\omega_i-\omega_f-\omega)^2+\frac{1}{4}(\Gamma_i+\Gamma_f)^2}$ for each transition between two lossy states $i\to f$, and exemplified in red in \autoref{fig:1}(c) for $\Gamma_i=\Gamma_f=\Gamma_p$  and $\omega_i-\omega_f=0.25$ eV.
The BR approach misses this effect (\autoref{fig:2}(c)), evaluating the transfer rate $K_{UP\to LP}$ only through the spectral density at the energy difference between the states associated with the maximum of $T_{\mathrm{UP\to LP}}(\omega)$ (e.g., at the vertical dashed blue line in \autoref{fig:1}(c)). Then, $K^{BR}_{\mathrm{UP\to LP}}=2\pi J_b(2g_{ec}) |\tilde{V}_{\mathrm{UP\to LP}}|^2$~\cite{del2015quantum} where $\tilde{V}_{\mathrm{UP\to LP}}=$(LP$|\hat{V}_s|$UP).
~In contrast, \BRLS{} captures the broadening of the states (\autoref{fig:2}(b)), and for each transition $i\to f$ naturally yields an effective spectral density that can be written as
\begin{equation}
J_{i\to f}^{\mathrm{eff}}(\omega_{i}-\omega_{f})=\int_0^\infty d\omega J_b(\omega)T_{i\to f}(\omega),\label{Jeff}
\end{equation}
such that $K_{\mathrm{UP\to LP}}^{\mathrm{BR\,LS}}=2\pi J_{\mathrm{UP\to
LP}}^{\mathrm{eff}}(2g_{ec})|\tilde{V}_{\mathrm{UP\to LP}}|^2$ (see SM).
Interestingly, $J_{\mathrm{UP\to LP}}^{\mathrm{eff}}(2g_{ec})$ is a convolution
of $J_b(\omega)$ with $T_{\mathrm{UP\to LP}}(\omega)$, highlighting the role of
the system in mediating the transfer of Markovianity between the two baths. We
emphasize that \autoref{Jeff} describes effective spectral densities for the system's interaction with the non-Markovian bath, which are
different for each system transition $i\to f$, in contrast to the case
of hierarchical environments~\cite{ma2014crossover, fruchtman2015quantum,
man2015non} where the baths directly interact with each other, and their combined effect can be described by a single spectral density including solely
their properties.


Finally, we analyze whether the \BRLS{} can be reproduced with the standard BR by replacing the spectral density $J_b(\omega)$ with a suitably broadened effective density. To do so, we consider the structured spectral density J$_0(\omega)$, plotted in black in \autoref{fig:3}, which has been taken from Ref.~\cite{zhao2020impact} and scaled down by two orders of magnitude to ensure weak coupling between nuclear and electronic degrees of freedom, as appropriate for J-aggregates~\cite{spano2015optical}. Moreover, we consider $N=30$ QEs in the system described by \autoref{Hm}, such that the system has dark states in addition to UP and LP\@. The other parameters are the same as for the test case above. 
Three transitions can occur in this setup: UP$\to$LP, UP$\to$DS, and DS$\to$LP\@. Since $T_{\mathrm{UP\to DS}} = T_{\mathrm{DS\to LP}}$, the last two transitions correspond to the same effective spectral density, J$_1^{\mathrm{eff}}(\omega)$, plotted in orange in \autoref{fig:3}. However, the transition UP$\to$LP that features a broader $T_{\mathrm{UP\to LP}}$ corresponds to another and broader effective spectral density, J$_2^{\mathrm{eff}}(\omega)$, plotted in light blue in \autoref{fig:3}.
As can be seen, both J$_1^{\mathrm{eff}}(\omega)$ and J$_2^{\mathrm{eff}}(\omega)$ are much smoother than J$_0(\omega)$, again indicating that a non-Markovian bath can become Markovian when acting on a lossy system.

\begin{figure}[t]
\includegraphics[width=\columnwidth, angle=0,scale=1,
draft=false,clip=true,keepaspectratio=true,valign=t]{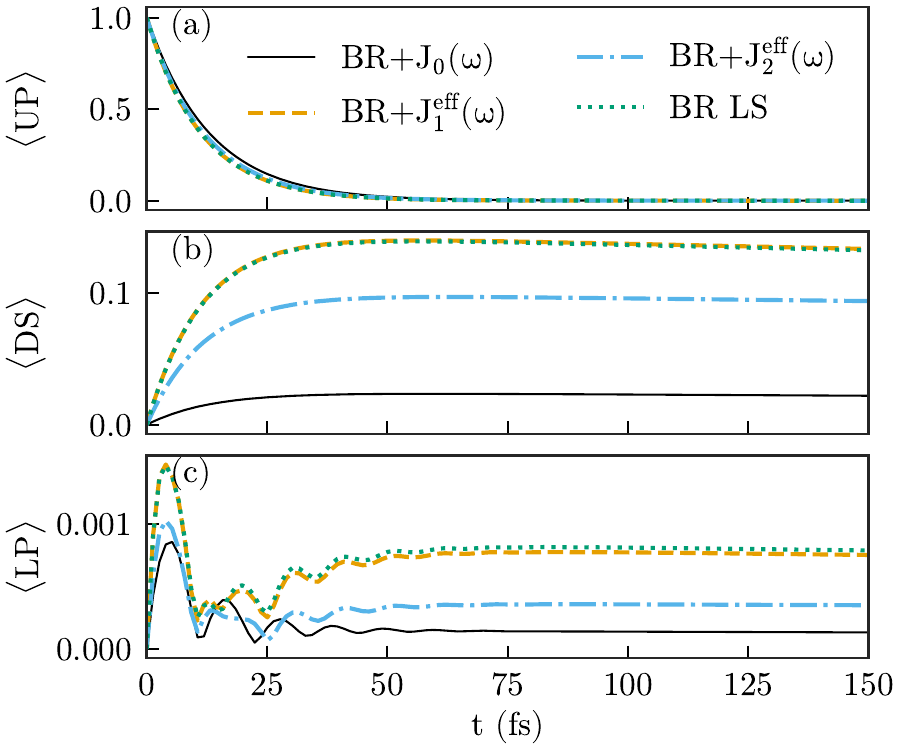}
\caption[]{The population of (a) UP,  (b) DS, and (c) LP, obtained by different approaches, as a function of time, when $30$ QEs interacting with the vibrational bath shown in black in \autoref{fig:3}, are strongly coupled to a lossy cavity ($g_{ec}=0.2$ eV). Initially, only the UP is populated. 
}
\label{fig:4}
\end{figure}

\autoref{fig:4} presents the population of the UP, DS and LP obtained by BR using J$_0(\omega)$, J$_1^{\mathrm{eff}}(\omega)$ and J$_2^{\mathrm{eff}}(\omega)$ and by \BRLS{} for $g_{ec}=0.2$~eV, when the UP is initially populated. Due to the complexity of the system, its exact solution is untractable.
The vertical lines in \autoref{fig:3} present the energy differences between the states, i.e., $\omega_{\mathrm{UP}}-\omega_{\mathrm{DS}}=\omega_{\mathrm{DS}}-\omega_{\mathrm{LP}}\approx g_{ec}$ and $\omega_{\mathrm{UP}}-\omega_{\mathrm{LP}}\approx 2g_{ec}$. As can be seen, the spectral densities are larger for the UP$\to$DS transition than for the UP$\to$LP one. As a result, and due to the presence of many DS, the dynamics in \autoref{fig:4} is dominated by the UP$\to$DS transition, followed by the DS$\to$LP transition.
Accordingly, the results obtained by BR$+$J$_1^{\mathrm{eff}}(\omega)$, associated with these transitions, are in excellent agreement with the \BRLS{} results, confirming that we can extract a linewidth-dependent effective spectral density from the \BRLS{} approach to describe the system's interaction with the non-Markovian bath. 
However, the results obtained by BR$+$J$_2^{\mathrm{eff}}(\omega)$ as well as by  BR$+$J$_0(\omega)$, which underestimates the spectral density at $\omega=g_{ec}$, present a smaller population transfer to the DS and therefore to the LP as well.
Note that the slight difference in the population of the LP between \BRLS{} and BR$+$J$_1^{\mathrm{eff}}(\omega)$ originates from the contribution of the UP$\to$LP transition, underestimated by J$_1^{\mathrm{eff}}(\omega)$ since J$_1^{\mathrm{eff}}(2g_{ec})<\mathrm{J}_2^{\mathrm{eff}}(2g_{ec})$. Moreover, the oscillations appearing in \autoref{fig:4}(c) are caused by the bath-induced Lamb shift of the system's energy levels~\cite{lamb1947fine}.

To conclude, we have explored a system interacting with two independent baths -- one Markovian and one non-Markovian. We identified the conditions under which the interaction with the non-Markovian bath becomes memoryless, a result of the interplay between the baths. Interestingly, this Markovian interaction stems from the dissipative nature the system acquires through its interaction with the Markovian bath, challenging the understanding that Markovianity is determined solely by the bath's properties.
To facilitate the computation of such interactions, we have extended the standard Bloch-Redfield (BR) theory, traditionally designed for Markovian baths, and derived the \BRLS{} approach, which assumes Markovian interaction due to the system losses rather than the bath's properties. This novel method efficiently computes the dynamics of a lossy system weakly coupled to a non-Markovian bath, especially in regimes where standard BR fails and more complex numerical methods~\cite{del2018tensor,prior2010efficient,strathearn2018efficient} would typically be required.
Given the prevalence of lossy systems in various fields, including optics~\cite{medina2021few}, acoustics~\cite{kitzman2023phononic}, quantum thermodynamics~\cite{brandner2017universal,tajima2021superconducting,zhang2022dynamical}, and chemistry~\cite{landau2017ab,gelbwaser2018high,jech2021quantum}, we expect our findings to improve understanding and facilitate less demanding calculations for a wide range of complex system-bath setups.

\begin{acknowledgments}
This work has been funded by the Spanish Ministry of Science, Innovation and Universities-Agencia Estatal de Investigación through grants PID2021-126964OB-I00, PID2021-125894NB-I00, EUR2023-143478, and CEX2018-000805-M (through the María de Maeztu program for Units of Excellence in R\&D). We also acknowledge financial support from the Proyecto Sin\'ergico CAM 2020 Y2020/TCS-6545 (NanoQuCo-CM) of the Community of Madrid, and from the European Union's Horizon Europe Research and Innovation Programme through agreement 101070700 (MIRAQLS). In addition, this project received funding from the European Union's Horizon 2020 research and innovation programme under the Marie Skłodowska-Curie Grant Agreement No 101034324.
\end{acknowledgments}

%

\end{document}


\title{Supplemental material for: Memory loss is contagious in open quantum systems}
\author{Anael Ben-Asher}
\affiliation{Departamento de Física Teórica de la Materia Condensada and Condensed Matter Physics Center (IFIMAC), Universidad Autónoma de Madrid, E28049 Madrid, Spain}

\author{Antonio I. Fernández-Domínguez}
\affiliation{Departamento de Física Teórica de la Materia Condensada and Condensed Matter Physics Center (IFIMAC), Universidad Autónoma de Madrid, E28049 Madrid, Spain}

\author{Johannes Feist}
\affiliation{Departamento de Física Teórica de la Materia Condensada and Condensed Matter Physics Center (IFIMAC), Universidad Autónoma de Madrid, E28049 Madrid, Spain}

\maketitle

\section{The derivation of the BR LS approach}
The Master equation describing the evolution of the density matrix $\rho_{s+b}(t)$ that encompasses the degrees of freedom of both a lossy system and an external bath is expressed as follows:
\begin{eqnarray}
 \frac{d{\rho_{s+b}}(t)}{dt}=-\frac{i}{\hbar}[\hat{H}_s+\hat{H}_b+\hat{H}_{sb}, \rho_{s+b}(t)]+\sum_A\gamma_A\hat{L}_{\hat{A}}\rho_{s+b}(t).\label{LinMEQ}
\end{eqnarray}
Here, $\hat{H}_s,\hat{H}_b$ represent the Hamiltonians of the system and bath, respectively. The interaction system-bath term, $\hat{H}_{sb}$, is given by $\hat{V}_s\otimes\int_0^\infty d\omega \sqrt{J(\omega)}(b_\omega+b_\omega^\dagger)$, where $\hat{V}_s$ is the system coupling operator, $J(\omega)$ is the bath's spectral density and $b_\omega$ ($b_\omega^\dagger$) is the bath bosonic annihilation (creation) operators. Additionally, $\hat{L}_{\hat{A}}\rho(t)$ is the Lindblad superoperator, given by $\hat{A}\rho(t) \hat{A}^\dagger-\frac{1}{2}\{\hat{A}^\dagger \hat{A},\rho(t)\}$, for the system jump operator $\hat{A}$ with the rate $\gamma_A$~\cite{breuer2002theory}.

First, we transform Eq.~\ref{LinMEQ} into the system-bath interaction picture that enables treating the bath perturbatively through the second-order Born approximation. To achieve this, we omit the terms $\hat{A}\rho(t) \hat{A}^\dagger$ in the Lindblad superoperator, responsible for incoherent transitions. This results in the equation:
$ \frac{d{\rho_{s+b}}(t)}{dt} = -\frac{i}{\hbar}[\hat{H}_{s+b}^{\mathrm{NH}}\rho_{s+b}(t)-\rho_{s+b}(t)\hat{H}_{s+b}^{\mathrm{NH}\dagger}]$, where $\hat{H}_{s+b}^{\mathrm{NH}}=\hat{H}_{s}^{\mathrm{NH}}+\hat{H}_b+\hat{H}_{sb}$ and $\hat{H}_{s}^{\mathrm{NH}}=\hat{H}_s-\frac{i}{2}\sum_A\gamma_A \hat{A}^\dagger \hat{A}$ is the effective non-Hermitian (NH) Hamiltonian describing the system dynamics~\cite{visser1995solution}. The transformation to the interaction picture is then achieved using the non-unitary operator $U(t)=e^{i(\hat{H}_s^{\mathrm{NH}}+\hat{H}_b)t/\hbar}$, yielding a master equation for the density matrix in the interaction picture $\rho_{s+b}^I(t)=U(t){\rho_{s+b}}(t)U^\dagger(t)$:
\begin{eqnarray}
\frac{d\rho_{s+b}^I(t)}{dt}=-\frac{i}{\hbar}\bigg(\hat{H}_{sb}^I(t)\rho_{s+b}^I(t) - \rho_{s+b}^I(t) \hat{H}_{sb}^{I\dagger}(t)\bigg) \label{Req}
\end{eqnarray}
where $\hat{H}_{sb}^I(t)=e^{i(\hat{H}_s^{\mathrm{NH}}+\hat{H}_b)t/\hbar}\hat{H}_{sb}e^{-i(\hat{H}_s^{\mathrm{NH}}+\hat{H}_b)t/\hbar}$ and $\hat{H}_{sb}$ is Hermitian. 
Integrating Eq.~\ref{Req} and substituting the resulting $\rho_{s+b}^I(t)$ into it, we obtain:
\begin{eqnarray}
\frac{d\rho_{s+b}^I(t)}{dt}=
-\frac{i}{\hbar}\bigg(\hat{H}_{sb}^I(t)\rho_{s+b}^I(0) - \rho_{s+b}^I(0) \hat{H}_{sb}^{I\dagger}(t) \bigg)\\\nonumber
-\frac{1}{\hbar^2}\int_0^t dt' \bigg(\hat{H}_{sb}^I(t) \hat{H}_{sb}^I(t') \rho_{s+b}^I(t')
- \hat{H}_{sb}^I(t) \rho_{s+b}^I(t') \hat{H}_{sb}^{I\dagger}(t') -\hat{H}_{sb}^I(t') \rho_{s+b}^I(t') \hat{H}_{sb}^{I\dagger}(t) + \rho_{s+b}^I(t') \hat{H}_{sb}^{I\dagger}(t') \hat{H}_{sb}^{I\dagger}(t)\bigg).
\end{eqnarray}

Next, we assume that the interaction is perturbative, such that $\rho_{s+b}(t)\approx\rho_s(t)\otimes\rho_b$ and $\rho_b$ is the time-independent and thermally populated density matrix of the bath. Tracing the bath's degrees of freedom results in a master equation for the system density matrix in the interaction picture ${\rho}_{s}^I(t)=\Tr_{B}[\rho_{s+b}^I(t)] = e^{i\hat{H}_s^{\mathrm{NH}}t/\hbar}{\rho}_s(t)e^{-i\hat{H}_s^{\mathrm{NH}\dagger} t/\hbar}$, given by:
\begin{eqnarray}\label{RedEq}
\frac{d{\rho}^I_{s}(t)}{dt}=
-\frac{1}{\hbar^2}\int_0^t dt' \bigg[C(t-t') \bigg(\hat{V}_{s}^I(t)\hat{V}_{s}^I(t'){\rho}_{s}^I(t') - \hat{V}_{s}^I(t') {\rho}_{s}^I(t') \hat{V}_{s}^{I\dagger}(t)\bigg) \\\nonumber
+ C^*(t-t')\bigg({\rho}^I_{s}(t') \hat{V}_{s}^{I\dagger}(t') \hat{V}_{s}^{I\dagger}(t)
 - \hat{V}_{s}^I(t) {\rho}^I_{s}(t') \hat{V}_{s}^{I\dagger}(t')\bigg) \bigg].
\end{eqnarray}
Here, $\hat{V}_s^I(t)=e^{i\hat{H}_s^{\mathrm{NH}}t/\hbar}\hat{V}_s e^{-i\hat{H}_s^{\mathrm{NH}}t/\hbar}$ and $ C(\tau)=\int_0^\infty d\omega J(\omega)\{(n(\omega)+1)e^{-i\omega\tau}+n(\omega)e^{i\omega\tau}\}$ is the autocorrelation function of a thermal bosonic bath with the spectral density $J(\omega)$ where $n(\omega)$ is the Bose occupation number. 

The BR LS approach is based on the assumption that the system's loss of coherence is responsible for the Markovianity of the interaction with the bath. Since the interaction picture used above does not capture the system loss, we cannot use it when assuming the Markovian approximation. Therefore, we apply on Eq.~\ref{RedEq} the inverse transformation to retrieve the  Lindblad terms:
\begin{eqnarray}
\frac{d\rho_{s}(t)}{dt}=
-\frac{i}{\hbar}[\hat{H}_s, \rho_s(t)]+\sum_A\gamma_A\hat{L}_{\hat{A}}\rho_s(t)\\\nonumber
-\frac{1}{\hbar^2}\int_0^t d\tau \bigg[C(\tau)\bigg(\hat{V}_s e^{-i\hat{H}_{s}^{\mathrm{NH}}\tau/\hbar}
\hat{V}_s\rho_{s}(t-\tau)e^{i\hat{H}_{s}^{\mathrm{NH}\dagger}\tau/\hbar} -e^{-i\hat{H}_{s}^{\mathrm{NH}}\tau/\hbar}\hat{V}_{s}\rho_{s}(t-\tau)e^{i\hat{H}_{s}^{\mathrm{NH}\dagger}\tau/\hbar}\hat{V}_{s}\bigg)\\ \nonumber+ C^*(\tau)\bigg(e^{-i\hat{H}_{s}^{\mathrm{NH}}\tau/\hbar}\rho_{s}(t-\tau)\hat{V}_{s}e^{i\hat{H}_{s}^{\mathrm{NH}\dagger}\tau/\hbar}\hat{V}_{s}
-\hat{V}_{s}e^{-i\hat{H}_{s}^{\mathrm{NH}}\tau/\hbar}\rho_{s}(t-\tau)\hat{V}_{s}e^{i\hat{H}_{s}^{\mathrm{NH}\dagger}\tau/\hbar}\bigg) \bigg]\nonumber.
\end{eqnarray}
Then, to remove the operators in the exponents, we use the identity operators described by the right and left eigenfunctions of $\hat{H}_s^{\mathrm{NH}}$ and $\hat{H}_s^{{NH}^\dagger}$, $\hat{I}=\sum_a|a)(a|$ and $\hat{I}=\sum_a|a^*)(a^*|$, where $\hat{H}_s^{\mathrm{NH}}|a)=(\hbar\omega_a-i\frac{\Gamma_a}{2})|a)$, 
$(a|\hat{H}_s^{\mathrm{NH}}=(\hbar\omega_a-i\frac{\Gamma_a}{2})(a|$, $\hat{H}_s^{\mathrm{NH}\dagger}|a^*)=(\hbar\omega_a+i\frac{\Gamma_a}{2})|a^*)$, and $(a^*|\hat{H}_s^{\mathrm{NH}\dagger}=(\hbar\omega_a+i\frac{\Gamma_a}{2})(a^*|$. Note that for eigenstates of a NH Hamiltonian, the notation $|\dots)$, $(\dots|$ rather than $|\dots\rangle$, $\langle\dots|$ is used to describe the right and left eigenstates~\cite{moiseyev2011non}.
In addition, we define $(a|\hat{V}_s|b)= \tilde{V}_{ab}$ and $(a^*|\hat{V}_{s}|b^*)= \tilde{V}_{ab}^*$ since $\hat{V}_s$ is a real and Hermitian operator. 
Consequently,
\begin{eqnarray}\label{s6}
\frac{d\rho_{s}(t)}{dt}=
-\frac{i}{\hbar}[\hat{H}_s, \rho_s(t)]+\sum_A\gamma_A\hat{L}_{\hat{A}}\rho_s(t)\\\nonumber
-\frac{1}{\hbar^2}\sum_{p,q,r,s}\int_0^t d\tau\bigg[C(\tau)\bigg(\tilde{V}_{pq} e^{-i(\hbar\omega_q-\frac{i}{2}\Gamma_q)\tau/\hbar}
\tilde{V}_{qr}|p)(r|\rho_{s}(t-\tau)|s^*)(s^*|e^{i(\hbar\omega_s+\frac{i}{2}\Gamma_s)\tau/\hbar}\\\nonumber -e^{-i(\hbar\omega_q-\frac{i}{2}\Gamma_q)\tau/\hbar}\tilde{V}_{qp}|q)(p|\rho_{s}(t-\tau)|s^*)(r^*|e^{i(\hbar\omega_s+\frac{i}{2}{\Gamma}_s)\tau/\hbar}\tilde{V}_{sr}^*\bigg)\\ \nonumber+ C^*(\tau)\bigg(e^{-i(\hbar\omega_s-\frac{i}{2}{\Gamma}_s)\tau/\hbar}|s)(s|\rho_{s}(t-\tau)|r^*)(p^*|\tilde{V}_{rq}^*e^{i(\hbar\omega_q+\frac{i}{2}{\Gamma}_q)\tau/\hbar}\tilde{V}_{qp}^*\\\nonumber
-\tilde{V}_{rs}e^{-i(\hbar\omega_s-\frac{i}{2}{\Gamma}_s)\tau/\hbar}|r)(s|\rho_{s}(t-\tau)|p^*)(q^*|\tilde{V}_{pq}^*e^{i(\hbar\omega_q+\frac{i}{2}{\Gamma}_q)\tau/\hbar}\bigg) \bigg]\nonumber.
\end{eqnarray}
One may notice that each term inside the integral in Eq.~\ref{s6} includes the decaying exponent $e^{-\frac{1}{2}(\Gamma_q+\Gamma_s)\tau/\hbar}$, which diminishes when $\tau\to\infty$. This decaying exponent encodes the impact of the system loss on its interaction with the bath and facilitates the Markovian assumption within the derivation of the BR LS approach for $t>\tau_{qs}=\frac{2}{\Gamma_q+\Gamma_s}$.
We use the coherent frame of the eigenenergies' difference to avoid artifacts originating from the loss of the memory of the internal dynamics of the system and assume $(i|\rho_s(t-\tau)|j^*)\approx(i|\rho_s(t)|j^*)e^{i(\omega_i-\omega_j)\tau}$. Moreover, we expand the upper limit of the integral to $\infty$.  
Subsequently, Eq.~\ref{s6} becomes Markovian and reads as:
\begin{eqnarray}\label{s7}
\frac{d\rho_{s}(t)}{dt}=
-\frac{i}{\hbar}[\hat{H}_s, \rho_s(t)]+\sum_A\gamma_A\hat{L}_{\hat{A}}\rho_s(t)\\\nonumber
-\frac{1}{\hbar^2}\sum_{p,q,r,s}\int_0^\infty d\tau\bigg[C(\tau)\bigg(\tilde{V}_{pq} 
\tilde{V}_{qr}|p)(r|\rho_{s}(t)|s^*)(s^*|e^{i(\omega_r-\omega_q)\tau}e^{-\frac{1}{2}(\Gamma_s+\Gamma_q))\tau/\hbar}\\\nonumber
-\tilde{V}_{qp}\tilde{V}_{sr}^*|q)(p|\rho_{s}(t)|s^*)(r^*|e^{i(\omega_p-\omega_q)\tau}e^{-\frac{1}{2}(\Gamma_s+\Gamma_q))\tau/\hbar}\bigg)\\ \nonumber+ C^*(\tau)\bigg(\tilde{V}_{rq}^*\tilde{V}_{qp}^*|s)(s|\rho_{s}(t)|r^*)(p^*|e^{-i(\omega_r-\omega_q)\tau}e^{-\frac{1}{2}(\Gamma_s+\Gamma_q))\tau/\hbar}\\\nonumber
-\tilde{V}_{rs}\tilde{V}_{pq}^*|r)(s|\rho_{s}(t)|p^*)(q^*|^{-i(\omega_r-\omega_q)\tau}e^{-\frac{1}{2}(\Gamma_s+\Gamma_q))\tau/\hbar}\bigg) \bigg]\nonumber.
\end{eqnarray}

Finally, we define $F_{jsq}=\int_0^\infty d\tau C(\tau)e^{-i(\omega_q-\omega_j)\tau}e^{-\frac{1}{2}(\Gamma_q+\Gamma_s)\tau/\hbar}$, and substitute $C(\tau)$ in it:
\begin{eqnarray}
{F}_{jsq}=\int_0^\infty \int_0^\infty d\tau   d\omega  e^{-\frac{1}{2}(\Gamma_q+\Gamma_s)\tau/\hbar} J(\omega)\bigg[(n(\omega)+1)e^{-i(\omega_q-\omega_j+\omega)\tau}+n(\omega)e^{-i(\omega_q-\omega_j-\omega)\tau}\bigg]=\\\nonumber
\int_0^\infty d\omega J(\omega)\bigg[(n(\omega)+1)\frac{\frac{1}{2\hbar}(\Gamma_q+\Gamma_s)-i(\omega_q-\omega_j+\omega)}{(\omega_q-\omega_j+\omega)^2+\frac{1}{4\hbar^2}(\Gamma_q+\Gamma_s)^2}
+n(\omega)\frac{\frac{1}{2\hbar}(\Gamma_q+\Gamma_s)-i(\omega_q-\omega_j-\omega)}{(\omega_q-\omega_j-\omega)^2+\frac{1}{4\hbar^2}(\Gamma_q+\Gamma_s)^2}
\bigg].
\end{eqnarray}
This simplifies Eq.~\ref{s7}, leading to the final Master equation within the BR LS approach, given by:
\begin{eqnarray}\label{lab}
\frac{d\rho_{s}(t)}{dt}=
-\frac{i}{\hbar}[\hat{H}_s, \rho_s(t)]+\sum_A\gamma_A\hat{L}_{\hat{A}}\rho_s(t)\\\nonumber
-\frac{1}{\hbar^2}\sum_{p,q,r,s}\bigg[F_{rsq}\tilde{V}_{pq} 
\tilde{V}_{qr}|p)(r|\rho_{s}(t)|s^*)(s^*| -F_{psq}\tilde{V}_{qp}\tilde{V}_{sr}^*|q)(p|\rho_{s}(t)|s^*)(r^*|\\ \nonumber+ F_{rsq}^*\tilde{V}_{rq}^*\tilde{V}_{qp}^*|s)(s|\rho_{s}(t)|r^*)(p^*|
-F_{psq}^*\tilde{V}_{rs}\tilde{V}_{pq}^*|r)(s|\rho_{s}(t)|p^*)(q^*|\bigg]\nonumber,
\end{eqnarray}
and can be rewritten by left multiplication with $(a|$ and right multiplication with $|b^*)$:
\begin{eqnarray}\label{BRLStensor}
\frac{d(a|\rho_{s}(t)|b^*)}{dt}=\\\nonumber-{i}\bigg(\omega_a-\omega_b-i\frac{\Gamma_a+\Gamma_b}{2\hbar}\bigg)(a|\rho_{s}(t)|b^*)
+\sum_{c,d}\bigg(\sum_A\gamma_A(a|\hat{A}|c)(d^*|\hat{A}^\dagger|b^*)-\frac{1}{\hbar^2}R_{a,b,c,d}\bigg)(c|\rho_{s}(t)|d^*),
\end{eqnarray}
where
$R_{a,b,c,d}=\delta_{bd}\sum_qF_{cdq}\tilde{V}_{aq}\tilde{V}_{qc}
+\delta_{ac}\sum_qF_{dcq}^*\tilde{V}^*_{dq}\tilde{V}_{qb}^*-(F_{cda}+F_{dcb}^*)\tilde{V}_{ac}\tilde{V}_{db}^*$ is the BR LS tensor used to solve the system dynamics.





\section{Extracting an effective spectral density from BR LS}
The BR LS results appearing in the figures in the main text are obtained without any further approximations to Eq.~\ref{BRLStensor}. 
However, to extract an effective spectral density from BR LS, we utilize the secular approximation where $(n|\rho_s(t)|m^*)=\delta_{nm}(n|\rho_s(t)|n^*)$.  
Under this approximation, we substitute $a=b=f$ and $c=d=i$ in Eq.~\ref{BRLStensor}. Moreover, we assume that the imaginary part of $\tilde{V}_{if}$ is negligible. This obtains:
\begin{eqnarray}
\frac{d(f|\rho_{s}(t)|f^*)}{dt}=-\bigg(\frac{\Gamma_f}{\hbar}+\sum_q(1-\delta_{qf})\frac{2Re[F_{ff q}]|\tilde{V}_{f q}|^2}{\hbar^2}\bigg)(f|\rho_{s}(t)|f^*)\\\nonumber
+\sum_{i}\bigg(\sum_A\gamma_A(f|A|i)(i^*|A^\dagger|f^*)+(1-\delta_{if})\frac{2Re[F_{iif}]|\tilde{V}_{if}|^2}{\hbar^2}\bigg)(i|\rho_{s}(t)|i^*).
\end{eqnarray}
We notice that the rate constant for the population transfer from state $i$ to state $f$ is given by 
\begin{eqnarray}
    K_{i\to f}^{BR LS} = \frac{2Re[F_{iif}] |\tilde{V}_{if}|^2}{\hbar} = 2\pi\int_0^\infty d\omega J(\omega)\bigg[(n(\omega)+1)T_{i\to f}(\omega)
+n(\omega)T_{i\to f}(-\omega)
\bigg]|\tilde{V}_{if}|^2\label{Kab},
\end{eqnarray}
where $T_{i\to f}(\omega)=\frac{1}{\pi}\frac{\frac{1}{2}(\Gamma_i+\Gamma_f)}{\hbar^2(\omega_f-\omega_i+\omega)^2+\frac{1}{4}(\Gamma_f+\Gamma_i)^2}$ is the spectrum of the transition $i\to f$, broadened due to the losses. By comparing $K_{i\to f}^{BR LS}$ with the constant rate obtained by BR, $K^{BR}_{i\to f}=2\pi J(\omega_i-\omega_f) |\tilde{V}_{if}|^2$~\cite{del2015quantum} when $n(\omega)=0$, we extract an effective spectral density for each transition $i\to f$,  given in Eq.~7 in the main text. 

\section{Benchamarking the performance of BR LS in obtaining the elements in the system's density matrix}

\begin{figure}[ht]
\begin{center}
   \includegraphics[width=0.5\columnwidth, angle=0,scale=1,
draft=false,clip=true,keepaspectratio=true]{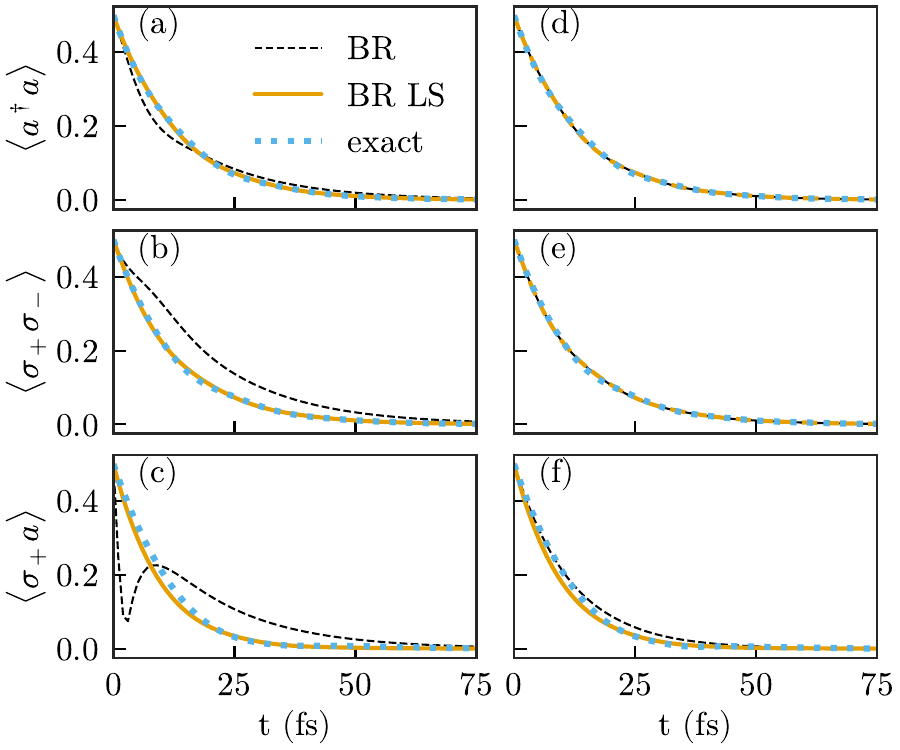}
\caption[]{The density matrix elements obtained by the standard BR approach, the introduced \BRLS{} approach, and numerically exact calculation for the test case studied in Fig.~2 in the main text. (a)--(c) show the results  when $g_{ec}=0.1$ eV, and (d)--(f) when $g_{ec}=0.11$ eV.
}
\label{fig:s1}
\end{center}
\end{figure}

In the main text, we demonstrate that the bath-driven population transfer can be accurately reproduced by the \BRLS{} approach when a lossy system weakly interacts with a non-Markovian bath. In \autoref{fig:s1}, we show that \BRLS{} also accurately reproduces the elements of the density matrix: $\langle a^\dagger a \rangle$, $\langle \sigma_+ \sigma_- \rangle$ and $\langle \sigma_+ a \rangle$, thus capturing the complete dynamics of the system. \autoref{fig:s1} presents the results obtained by the BR, the \BRLS{}, and the numerically exact approaches for the test case given in Fig.~2 in the main text. In (a)--(c), the results for $g_{ec}=0.1$ eV are presented, and in (d)--(f) for $g_{ec}=0.11$ eV. All the other parameters are the same as for Fig.~2 in the main text. As can be seen, in both cases, \BRLS{} is in excellent agreement with the numerically exact results and is capable of describing the exact dynamics. 

\section{Benchamarking the performance of BR LS when the NH Hamiltonian description is not valid}

\begin{figure}[ht!]
\begin{center}
   \includegraphics[width=0.5\columnwidth, angle=0,scale=1,
draft=false,clip=true,keepaspectratio=true]{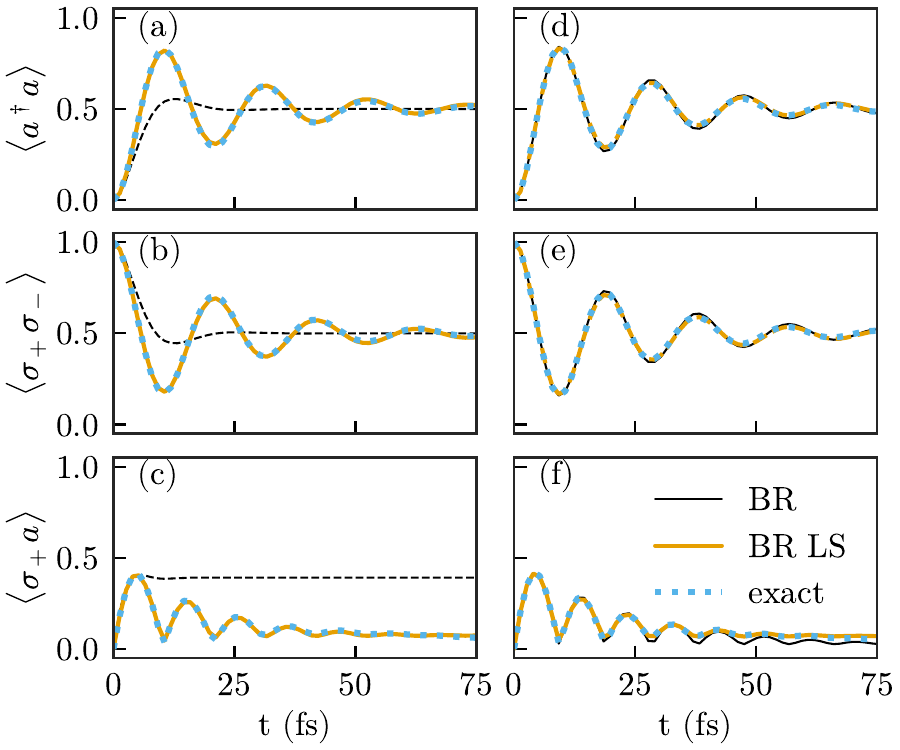}
\caption[]{The density matrix elements obtained by the standard BR approach, the introduced \BRLS{} approach, and numerically exact approach for the case where a NH Hamiltonian can not capture the complete dynamics. (a)--(c) show the results when $g_{ec}=0.1$ eV, and (d)--(f) when $g_{ec}=0.11$ eV.
}
\label{fig:s2}
\end{center}
\end{figure}

When the jump operators in the Lindblad terms represent dephasing rather than decay, the NH Hamiltonian, which only describes the decay process, cannot be applied to describe the dynamics and to capture the repopulation by the feeding terms. However, we show that it can still be used to describe the broadening of the system’s energy levels and, therefore, to build the BR LS tensor using the Markovian assumption. To do so, we study the test case described in the main text with a dephasing rate rather than decay rates. 
Namely, we consider solely the jump operator $\hat{A}=a^\dagger a$ in the Lindblad term with dephasing rate $\gamma_c=0.1$ eV, while the system's Hamiltonian is:
\begin{eqnarray}
\hat{H}_s = {\omega}_c a^\dagger a +{\omega}_c\sigma_+\sigma_-+g_{ec}(a^\dagger \sigma_-+\sigma_+ a). \label{Hh}
\end{eqnarray}
Since an effective NH Hamiltonian cannot predict the dynamics of such a setup, rather than studying the population transfer between the eigenstates of the NH Hamiltonian (e.g., the polaritons), we explore in \autoref{fig:s2} the performance of \BRLS{} in reproducing the elements of the density matrix as a function of time when the initial state is $\sigma_+|0\rangle$. For comparison, we also plot the results obtained by the BR and the numerically exact approaches. In (a)--(c), the results obtained for $g_{ec}=0.1$ eV are presented, and in (d)--(f) for $g_{ec}=0.11$ eV. 
The bath's parameters are the same as in the test case in the main text.  
As can be seen,  \BRLS{} is in excellent agreement with the numerically exact results. 
This shows that any interaction with a Markovian bath, even if yielding dephasing rather than decay, can acquire to the system a memoryless character that the NH Hamiltonian captures. As a result, the \BRLS{} is also valid for cases where the NH Hamiltonian fails to reproduce the complete system's dynamics.

\section{Isolation of the dynamics induced by the non-Markovian bath using the NH formalism}
\begin{figure}[ht!]
\begin{center}
   \includegraphics[width=0.5\columnwidth, angle=0,scale=1,
draft=false,clip=true,keepaspectratio=true]{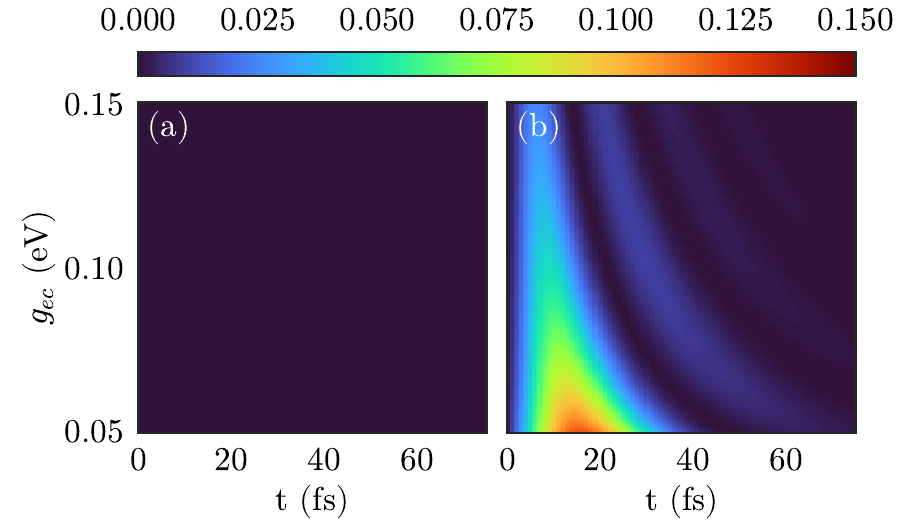}
\caption[]{The population transfer between the eigenstates of (a) the NH effective Hamiltonian, (b) the Hermitian Hamiltonian, when $g_b=0$, i.e., no interaction with the non-Markovian bath.
}
\label{fig:s4}
\end{center}
\end{figure}

To show that the population transfer studied in the main text indeed isolates the dynamics induced by the non-Markovian bath, in \autoref{fig:s4} (a), we present the population of the LP  when there is no coupling to the non-Markovian bath, i.e., when $g_b=0$. The other parameters are the same as for Fig.~2 in the main text. As can be seen, no population transfer is obtained for this case, showing that the populations of the eigenstates of the NH Hamiltonian (the polaritons) indeed isolate the effect of the non-Markovian bath and, therefore, facilitate studying its memoryless interaction with the system.
For comparison, in \autoref{fig:s4} (b), the population of the lower eigenstate in the first excitation manifold of the Hermitian Hamiltonian (the real part of Eq.~5 in the main text) is presented where the initial state is the upper eigenstate in the same manifold. The other parameters are the same as for \autoref{fig:s4} (a). As can be seen, since the system's Hermitian Hamiltonian does not include the information on the Markovian bath, the population transfer for no coupling with the non-Markovian bath is not zero but also consists of the dynamics induced by the former bath, showing the strength of the NH Hamiltonian in studying the interplay between the baths.

\section{Measuring the non-Markovianity using the trace distance method}
\begin{figure}[ht!]
\begin{center}
   \includegraphics[width=0.5\columnwidth, angle=0,scale=1,
draft=false,clip=true,keepaspectratio=true]{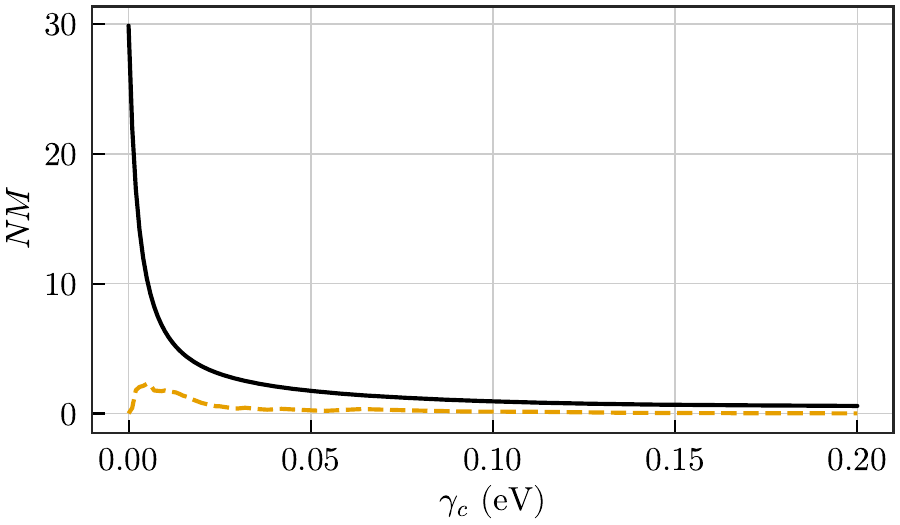}
\caption[]{The NM measured by the trace distance method. While the solid black line considers the whole time evolution, the dashed orange one takes into account just the dynamics occurring after the system's lifetime $\tau_s=\frac{1}{\gamma_c}$.
}
\label{fig:s3}
\end{center}
\end{figure}

In this section, we use the trace distance method~\cite{breuer2009measure}, designed for measuring the non-Markovianity in open quantum systems, to analyze the effects of the systems losses on the memoryless character of the dynamics in the test case studied in the main text. In this approach, the non-Markovianity is measured by 
\begin{eqnarray}\label{tr_di}
NM=\max_{\rho_1(0),\rho_2(0)}\sum_i[D(\rho_1(b_i),\rho_2(b_i))-D(\rho_1(a_i),\rho_2(a_i))]
\end{eqnarray}
where $D(\rho_1(t),\rho_2(t))$ is the trace distance between two density matrices of the system, and the index $i$ sums on all the time intervals $(a_i,b_i)$ for which the trace distance increases. 
Similar to Ref.~\cite{breuer2009measure}, we find the optimal pair of initial states for which the sum in Eq.~\ref{tr_di} is maximal by checking a sufficiently large sample of pairs of initial states. This yields that the optimal pair is $\rho_1(0)=|+\rangle\langle+|,\rho_2(0)=|-\rangle\langle-|$ where $|\pm\rangle=\frac{1}{\sqrt{2}}(\sigma_+\pm i a^\dagger)|0\rangle$.
In \autoref{fig:s3}, we plot the computed non-Markovianity, describing the backflow of information from the non-Markovian bath into the system, in black as a function of $\gamma_c$ (the other parameters are the same as for Fig.~2 in the main text). As can be seen, when the system loss, reflecting the coupling with the Markovian bath, increases, the non-Markovianity significantly drops. This confirms the transfer of Markovianity between the baths. 

To show that indeed the system losses, and not solely the baths' properties, are responsible for the memoryless dynamics, we plot in a dashed line in \autoref{fig:s3} the $NM$ calculated by Eq.~\ref{tr_di} only using the time intervals $(a_i,b_i)$ that obey $a_i>\tau_s=\frac{1}{\gamma_c}$. 
While for small values of $\gamma_c$, this quantity increases with $\gamma_c$ since it takes into account more time intervals, the trend changes when $\gamma_c>\kappa=5$ meV, i.e., when the system decays faster than the bath's autocorrelation function. In this regime,  the non-Markovianity goes to zero, showing that the dynamics become fully Markovian for times longer than the lifetime of the system. This is in accordance with the observed agreement between the \BRLS{} and the numerically exact method presented in Fig.~2 in the main text and demonstrates the role of the system losses in inducing Markovian dynamics.

%